\documentclass[12pt]{article}

\usepackage{amssymb}

\begin{document}

\begin{titlepage}

\begin{flushright}
%arXiv:YYMM.XXXX
\end{flushright}
\vskip 2.5cm

\begin{center}
{\Large \bf Laboratory Bounds on Electron Lorentz Violation}
\end{center}

\vspace{1ex}

\begin{center}
{\large Brett Altschul\footnote{{\tt baltschu@physics.sc.edu}}}

\vspace{5mm}
{\sl Department of Physics and Astronomy} \\
{\sl University of South Carolina} \\
{\sl Columbia, SC 29208} \\

\end{center}

\vspace{2.5ex}

\medskip

\centerline {\bf Abstract}

\bigskip

Violations of Lorentz boost symmetry in the electron and photon sectors can
be constrained by studying several different high-energy phenomenon. Although they
may not lead to the strongest bounds numerically, measurements made in
terrestrial laboratories produce the most reliable results. Laboratory bounds can
be based on observations of synchrotron radiation, as well as the observed
absences of vacuum Cerenkov radiation ($e^{\pm}\rightarrow e^{\pm}+\gamma$)
and photon decay ($\gamma\rightarrow e^{+}+e^{-}$). Using measurements of synchrotron
energy losses at LEP and the survival of TeV photons, we place new bounds on the
three electron Lorentz violation coefficients $c_{(TJ)}$, at the
$3\times 10^{-13}$ to $6\times 10^{-15}$ levels.

\bigskip

\end{titlepage}

\newpage

\section{Introduction}

Recent years have seen a surge of interest in the possibility that apparently
fundamental symmetries, such as Lorentz and CPT invariances, may not be exact in
nature. The scope and precision of experimental tests of such symmetry violations
have expanded tremendously, but no compelling experimental evidence for violation of
Lorentz invariance or CPT has been seen. Nonetheless, this remains a very active
field of research. One reason for this is that the payoff from any discovery would
be amazingly high; such symmetry violations (or other similar effects that could not
occur within the known contexts of the standard model and general relativity) would
be immediate evidence for new physics of a completely unheralded character. For
example, violations of Lorentz symmetry or CPT could be tied to the structure of
quantum gravity and would provide crucial clues as to what other novel effects we
should look for.

All possible forms of Lorentz and CPT violations consistent with quantum theory
as we understand it can be parameterized through an effective field theory called the
standard model extension (SME). The SME contains Lorentz- and CPT-violating
corrections to the standard model~\cite{ref-kost1,ref-kost2}. (The framework can
also be expanded to cover gravity~\cite{ref-kost12}.)
Both the renormalizability~\cite{ref-kost4,ref-colladay2} and
stability~\cite{ref-kost3} of the SME have been studied.
The SME is now the standard framework for parameterizing the results of experimental
Lorentz and CPT tests. With the systematic description made possible by the SME,
experimentalists can study a wider range of symmetry-breaking phenomena; most possible
forms of Lorentz and CPT violation were overlooked in earlier, purely phenomenalistic
analyses.

Recent searches for Lorentz violation have included studies of matter-antimatter
asymmetries for trapped charged
particles~\cite{ref-bluhm1,ref-gabirelse,ref-dehmelt1} and bound state
systems~\cite{ref-bluhm3,ref-phillips},
measurements of muon properties~\cite{ref-kost8,ref-hughes}, analyses of
the behavior of spin-polarized matter~\cite{ref-heckel3},
frequency standard comparisons~\cite{ref-berglund,ref-kost6,ref-bear,ref-wolf},
Michelson-Morley experiments with cryogenic
resonators~\cite{ref-muller3,ref-herrmann2,ref-herrmann3,ref-eisele,ref-muller2},
Doppler effect measurements~\cite{ref-saathoff,ref-lane1},
measurements of neutral meson
oscillations~\cite{ref-kost10,ref-kost7,ref-hsiung,ref-abe,
ref-link,ref-aubert}, polarization measurements on the light from cosmological
sources~\cite{ref-carroll2,ref-kost11,ref-kost21,ref-kost22},
high-energy astrophysical
tests~\cite{ref-stecker,ref-jacobson1,ref-altschul6,ref-altschul7,ref-klinkhamer2},
precision tests of gravity~\cite{ref-battat,ref-muller4}, and others.
The results of these experiments set constraints on the various SME coefficients, and
up-to-date information about
most of these constraints may be found in~\cite{ref-tables}.

Some forms of Lorentz violation---those which break boost invariance---may best be
constrained by using relativistic particles. When greater velocities are involved in a
test, more stringent bounds on certain SME coefficients are possible.
There are two important high-energy processes that are absolutely forbidden in the
absence of Lorentz violation---vacuum Cerenkov radiation and the photon decay
$\gamma\rightarrow e^{+}+e^{-}$. If Lorentz violation appears at high energies,
there may be thresholds above which these processes are allowed. The typical scale
of such thresholds is $\frac{m}{\sqrt{|\delta|}}$, where $m$ is the mass of the
charged particles involved and $|\delta|$ is the dimensionless scale of the Lorentz
violation. Consequently, the non-occurrence of these processes up to an energy $E$
will constrain the relevant forms of Lorentz violation at the
${\cal O}\left(\frac{m^{2}}{E^{2}}\right)$ level.

The fastest-moving particles we can study are astrophysical in origin, and
astrophysical observations have been used to place a number of very strong
bounds. However,
it is also desirable to have laboratory bounds on the SME coefficients. Bounds
that are based on energetic astrophysical phenomena may rely on our having an
accurate understanding of processes occurring very far away. In many cases, what
particle interactions are responsible for a particular experimentally observed
phenomenon may only be inferred, not determined directly; and these inferences may be
controversial. A relevant example is the
disagreement whether TeV $\gamma$-rays that reach the Earth are produced mostly by 
inverse Compton scattering or $\pi^{0}$ decay. Conclusions that are based on
inferences about astrophysical processes are useful; however,
they should also be used with some caution. Such conclusions may not be as reliable
as similar ones based on observations made directly in the laboratory.

We shall conclude this introductory
section by presenting the SME Lagrangian relevant for
interactions among electron, positrons, and photons at high energies. In
section~\ref{sec-kinematics} we shall discuss the how the velocity, momentum,
and other kinematic quantities are related in the Lorentz-violating theory.
Section~\ref{sec-vc} examines how the absence of vacuum Cerenkov radiation from
electron and positron beams at LEP, as well as the absence of photon decay, can be
used to
place bounds on SME parameters. In section~\ref{sec-sync}, we discuss the bounds
that can be placed by studying the actual radiation emission (which is synchrotron
in origin) from the LEP beams. The paper concludes with section~\ref{sec-comb},
which collects the best reliable bounds that can be derived from all the various
laboratory techniques and discusses the outlook for future improvements.

The electron and photon sectors of the minimal SME are described by the Lagrange
density
\begin{eqnarray}
{\cal L} & = & -\frac{1}{4}F^{\mu\nu}F_{\mu\nu}
-\frac{1}{4}k_{F}^{\mu\nu\rho\sigma}F_{\mu\nu}F_{\rho\sigma}
+\frac{1}{2}k_{AF}^{\mu}\epsilon_{\mu\nu\rho\sigma}F^{\nu\rho}A^{\sigma}
+\bar{\psi}(i\Gamma^{\mu}D_{\mu}-M)\psi \\
\Gamma^{\mu} & = & \gamma^{\mu}+c^{\nu\mu}\gamma_{\nu}-d^{\nu\mu}\gamma_{\nu}
\gamma_{5}+e^{\mu}+if^{\mu}\gamma_{5}+\frac{1}{2}g^{\lambda\nu\mu}
\sigma_{\lambda\nu} \\
M & = & m+\!\not\!a-\!\not\!b\gamma_{5}+\frac{1}{2}H^{\mu\nu}\sigma_{\mu\nu}+im_{5}
\gamma_{5}.
\end{eqnarray}
The unconventional behavior of electrons and photons at high energies is primarily
determined by the coefficients $c$, $d$, and $k_{F}$, which are dimensionless and
preserve CPT. The fermion coefficients $c$ and $d$ are traceless,
while $k_{F}$ has a vanishing double trace and exhibits the same symmetries as
the Riemann tensor.
These coefficients break rotation, boost, and parity symmetries, and
they modify the velocities of
electrons, positrons, and photons. However, the usual $ev^{\mu}A_{\mu}$
coupling between charged sources and the electromagnetic field is unchanged;
this is a consequence of electromagnetic gauge invariance.
Since non-gravitational Lorentz violation is necessarily a small effect, we shall
only consider the leading-order effects of the various SME coefficients; at that
order, only the symmetric parts of $c$ and $d$ can contribute to physical effects.

\section{Lorentz-Violating Kinematics}
\label{sec-kinematics}

The $c$, $d$, and $k_{F}$ coefficients affect the energy-momentum relations for
relativistic quanta. In the electron sector, the changes are most easily
expressed in terms of modifications to the maximum achievable velocity (MAV) of a
particle. The MAV in the direction $\hat{v}$ is~\cite{ref-altschul4}
\begin{equation}
1+\delta(\hat{v})=
1-c_{00}-c_{(0j)}\hat{v}_{j}-c_{jk}\hat{v}_{j}\hat{v}_{k}+sd_{00}+sd_{(0j)}
\hat{v}_{j}+sd_{jk}\hat{v}_{j}\hat{v}_{k},
\end{equation}
where $s$ the product
of a particle's helicity and fermion number ($+1$ for electrons, $-1$ for
positrons). The parentheses indicate a symmetrized sum, such as
$c_{(0j)}=c_{0j}+c_{j0}$.
The dispersion relation for a relativistic
particle is $E=\sqrt{m^{2}+p^{2}[1+2\delta(\hat{p})]}$; this determines the
threshold for various novel processes that depend on the existence of Lorentz
violation. The velocity at ultrarelativistic energies is
\begin{equation}
\vec{v}=\left[1+\delta(\hat{p})-\frac{m^{2}}{2E^{2}}\right]\hat{p},
\end{equation}
which is the usual velocity plus $\delta(\hat{p}\,)\hat{p}$.

The effects in the electromagnetic sector are similar. Photons, being massless,
always move
with their MAV. The spin-dependent part of $k_{F}$ is extremely tightly constrained
experimentally; it gives rise to birefringence that is not seen, even for photons
that have traversed cosmological distances. The non-birefringent part is
$\tilde{k}^{\mu\nu}=(k_{F})_{\alpha}\,^{\mu\alpha\nu}$, which changes the speed of
light in the direction $\hat{v}$ to $1+\delta_{\gamma}(\hat{v})=
1-\frac{1}{2}\left[\tilde{k}_{00}+
2\tilde{k}_{0j}\hat{v}_{j}+\tilde{k}_{jk}\hat{v}_{j}\hat{v}_{k}\right]$. Most
observable relativistic effects
depend on the difference between the velocity of a moving charge and the speed of
light in the same direction. For example, if an electron's speed exceeds
$1+\delta_{\gamma}(\hat{v})$, vacuum Cerenkov radiation will be emitted; this is
analogous to ordinary Cerenkov emission in matter, where the speed of light in a
material is less than 1.

We may choose coordinates in such a way as to eliminate the $c$ or $\tilde{k}$
coefficients in one sector of the theory. (Only differences between these
coefficients in different sectors are physically observable.)
We shall use this coordinate freedom to make the
photon sector conventional. All the constraints we derive will therefore be
formulated as limits on combinations of
electron coefficients. However, bounds on $c^{\nu\mu}$ should be reinterpreted as
bounds on the linear combinations $c^{\nu\mu}-\frac{1}{2}\tilde{k}^{\mu\nu}$ if
the electromagnetic $\tilde{k}$ coefficients are not set to zero.

Constraints on SME coefficients are conventionally stated in a Sun-centered
celestial equatorial coordinate frame~\cite{ref-bluhm4}.
The origin lies at the center of the Sun.
The $Z$-axis points along the Earth's rotation axis; the $X$-axis
points toward the vernal equinox point on the celestial sphere; and the
$Y$-axis is  set by the right hand rule. Time in these coordinates is $T$.

\section{Vacuum Cerenkov Radiation and Photon Decay}
%at Accelerators
\label{sec-vc}

We are interested in the effects of Lorentz violation on relativistic electron-photon
dynamics in terrestrial laboratories. The most energetic Earthbound electron and
positron beams were created at the Large Electron-Positron Collider (LEP).
%and the most energetic photons were at the Tevatron.
In order to study direction-dependent effects, we must understand the range of beam
velocities that were possible at this accelerator.
In the course of each revolution around a circular accelerator, a particle
moves in every direction that lies in the plane of the accelerator ring. As the Earth
rotates, this plane rotates as well. If the accelerator is locally horizontal, the
plane of the accelerator ring sweeps out the exterior of a double
cone. The accelerated particles will travel in every direction $\hat{v}$ that
satisfies $|\hat{v}\cdot\hat{Z}|\leq\sin\chi$ (where $\chi$ is the colatitude of the
laboratory) during the course of one sidereal day.
LEP was located at the European Organization for Nuclear Research (CERN), at latitude
46.2$^{\circ}$ N.
%and Fermilab (41.8$^{\circ}$ N). The Large Hadron
%Collider (LHC) has now exceed the Tevatron in operating energies; as the LHC moves
%to progressively higher energies and photon production data become available, there
%will be some improvement in the laboratory photon splitting bounds. However, the
%strength of many two-sided constraints is determined by the weaker
%LEP vacuum Cerenkov bounds, and the production of more energetic $\gamma$-rays at
%accelerators will provide only a modest improvement. Moreover, there is no reason
%why the observed absence of $\gamma\rightarrow e^{+}+e^{-}$ with laboratory-produced
%photons provides more reliable result than the similar survival of
%astrophysically-generated $\gamma$-rays. The survival of any photon of a given
%energy for an appreciable period of time gives a reliable constraint on the relevant
%forms of Lorentz violation.

It is immediately natural to ask what effect the Lorentz-violating threshold for
vacuum Cerenkov radiation would have had on LEP's operation.
The threshold for an electron or positron to emit such radiation
is~\cite{ref-altschul9}
\begin{equation}
\label{eq-ET}
E_{T}=\frac{m}{\sqrt{2\delta(\hat{v})}}.
\end{equation}
This is the energy at which the charge's speed reaches the speed of light. $E_{T}$
is a real energy only if $\delta(\hat{v})>0$; if $\delta(\hat{v})<0$, the MAV is
less than 1.

The instantaneous absence of vacuum Cerenkov radiation indicates
that
\begin{equation}
\label{eq-vcbound}
\delta(\hat{v})<A\equiv\frac{m^{2}}{2E^{2}},
\end{equation}
because the non-radiating particles cannot be moving superluminally. 
The Cerenkov bound supplies only a one-sided constraint on $c$; the absence
of photon decay can supply the other side. However, the vacuum Cerenkov bounds
constrain the $d$-dependent part of $\delta(\hat{v})$ from both sides. Both helicity
states were represented in the LEP beams, and if Cerenkov radiation from either
spin state were allowed, it would have been observed.
However, the bounds on $d$ turn out not to be that interesting. All nine of these
spin-dependent coefficients have already been
constrained using laboratory experiments with torsion pendulums~\cite{ref-heckel3}.
The pendulum bounds are substantially stronger than the bounds quoted here,
although they are
not on the $d$ coefficients individually but on particular linear combinations of
the $d$ and other parameters. However, it would require remarkable fine tuning
for the observable combinations to be as small as they are without
the $d$ coefficients also being well below the $10^{-11}$--$10^{-12}$ scale
accessible via accelerators.

Vacuum Cerenkov radiation is an extremely efficient form of energy loss.
If the maximum LEP beam energy of $E_{b}=104.5$ GeV were only 1\% above the Cerenkov
threshold $E_{T}$, the expected time required for a superluminal electron or positron
to emit
a $\gtrsim 1$ GeV photon and drop below the Cerenkov threshold would be
$\lesssim 10^{-9}$ s. During this time, the particle
would advance about 0.2 m along the
27 km track, and its direction would not change appreciably. The time required
for vacuum Cerenkov emission to drain away a substantial amount of the beam's energy
is so short that we may conclude that for every direction along the beam's path,
$E_{b}$ is no greater than $1.01E_{T}$. The right-hand-side of (\ref{eq-vcbound})
may therefore be taken to be
$A=\frac{m^{2}}{2(0.99E_{b})^{2}}=1.2\times10^{-11}$~\cite{ref-hohensee1}.

The survival of energetic photons (through the absence of
$\gamma\rightarrow e^{+}+e^{-}$) can also be used to place constraints on Lorentz
violation; if this process were allowed, it would operate extremely
rapidly~\cite{ref-coleman2,ref-hohensee2}.
The threshold for photon decay is $2|E_{T}|$, defined only if $E_{T}$ as given by
(\ref{eq-ET}) is imaginary. The absence of this process for photons of energy
$E_{\gamma}$ moving in the direction $\hat{v}$ implies
\begin{equation}
\label{eq-photonbound}
c_{00}+c_{(0j)}\hat{v}_{j}+c_{jk}\hat{v}_{j}\hat{v}_{k}<B\equiv\frac{2m^{2}}
{E_{\gamma}^{2}}.
\end{equation}
The bound (\ref{eq-photonbound}) does not involve $d$. The reason is that the
electron and positron produced by a single photon decaying at threshold have parallel
spins. Their energy-momentum relations are shifted in opposite directions by $d$; as
a consequence, the presence of $d$ will not make the otherwise forbidden process
allowed. This ``no-go'' conclusion can be
evaded by involving additional fields in the decay or taking into account angular
momentum nonconservation, which is second order in the Lorentz violation; however,
these cannot induce the kind of rapid photon decay that may be seen with
nonzero $c$.

Collision events at accelerators produce many photons. Photons were produced at the
Tevatron with energies above 300 GeV~\cite{ref-abazov}.
However, matters are complicated by the fact that the range of photon
directions is essentially the entire sphere. If isotropy is assumed, then the
observed survival of these photons may easily be converted into bounds on the SME
coefficient $c_{00}$; However, if anisotropy
is allowed, the analysis of the accelerator-produced photons becomes much more
complicated. Moreover, even with the simplification of assumed
rotation invariance, these
data do not provide a particularly efficient way to set reliable constraints on
Lorentz violation.
For one thing, the laboratory origin of the photons is not a requirement to produce a
reliable bound. Any photon, regardless of origin, that is observed to have propagated
a substantial distance is a demonstration that $\gamma\rightarrow e^{+}+e^{-}$ is
not allowed at the momentum involved. The detection of the photon, which is the
only part of its history that really matters, is just as much a reliable, laboratory
measurement as the observation of circulating beams at an accelerator.
Cosmic ray $\gamma$-rays have been observed
with energies of nearly 100 TeV, which should lead to stronger bounds than would
be possible with accelerator data alone.
%However, for purposes of determining the
%strongest reliable bounds on various forms of Lorentz violation, the photon survival
%data turn out to be superfluous; the strength of the final bounds is controlled by
%other factors.

We shall therefore not discuss the decay of photons produced at accelerators any
further, except to note the following.
The beam directions $\hat{v}$ explored at LEP are a subset of the photon
directions observed at the Tevatron. If the bounds (\ref{eq-vcbound}) and
(\ref{eq-photonbound}) were simultaneously
satisfied for every direction with $|\hat{v}\cdot\hat{Z}|\leq\sin\chi_{C}$
(where $\chi_{C}\approx43.8^{\circ}$ is the colatitude at CERN), then for any such
direction, we would
have the vacuum Cerenkov bounds $c_{TT}\pm c_{(TJ)}\hat{v}_{J}+c_{JK}\hat{v}_{J}
\hat{v}_{K}-|d_{TT}\pm d_{(TJ)}\hat{v}_{J}+d_{JK}\hat{v}_{J}\hat{v}_{K}|>-A$
and photon decay bounds $c_{TT}\pm c_{(TJ)}\hat{v}_{J}+c_{JK}\hat{v}_{J}
\hat{v}_{K}<B$. (The $\pm$ arises from taking into account the motion both along
$\hat{v}$ and $-\hat{v}$.)
It would then be straightforward to separate the $c$ and $d$ bounds and to
further disentangle the even- and odd-parity coefficients. Adding and subtracting
these inequalities yields
\begin{eqnarray}
\label{eq-cTTbound}
-A \,\, < \,\, c_{TT}+c_{JK}\hat{v}_{J}\hat{v}_{K} & < & B \\
|c_{(TJ)}\hat{v}_{J}| & < & (A+B)/2 \\
\label{eq-dTTbound}
|d_{TT}+d_{JK}\hat{v}_{J}\hat{v}_{K}| & < & A+B \\
\label{eq-dTJbound}
|d_{(TJ)}\hat{v}_{J}| & < & A+B.
\end{eqnarray}
These would be the maximal anisotropic generalizations of the LEP vacuum Cerenkov
radiation and Tevatron photon decay results of~\cite{ref-hohensee1}.
They would be sufficient to establish completely separate bounds on the six
$c_{(TJ)}$
and $d_{(TJ)}$ coefficients, by choosing $\hat{v}=\hat{X}$, $\hat{v}=\hat{Y}$ and
$\hat{v}=\hat{X}\cos\chi_{C}\pm\hat{Z}\sin\chi_{C}$. [The resulting bound on, for
example, $c_{(TZ)}$ would be $|c_{(TZ)}|<\frac{A+B}{2\sin\chi_{C}}$.]
However, the other inequalities (\ref{eq-cTTbound}) and (\ref{eq-dTTbound}) could
not be further disentangled using only this data. We shall therefore turn our
attention to other processes that can occur at accelerators.

%Other laboratory bounds come from experiments with electromagnetic oscillators. Such
%oscillators' resonant frequencies obviously depend on their shapes, which in turn
%depend on the structure of the electron sector.
%We note, however, that these tests may not be sensitive to precisely the same clean
%$c^{\nu\mu}-\frac{1}{2}\tilde{k}^{\mu\nu}$ parameters as the accelerator
%experiments.
%The oscillators tests involve comparisons with outside frequency standards, which
%general introduce dependences on hadronic Lorentz violation as well. If there is
%no Lorentz violation outside the electron and photon sectors, the inclusion of the
%laboratory bounds makes possible a further disentanglement of the coefficients
%involved in the accelerator tests.

\section{Synchrotron Radiation}
\label{sec-sync}

Vacuum Cerenkov radiation and photon decay are normally forbidden,
but they may occur in the presence of Lorentz violation. On the other hand,
the most important electron-photon interaction that we know really does occur at
particle accelerators is synchrotron radiation, and information about real
synchrotron losses can
provide further constraints on electron Lorentz violation. Information about
these losses provides the strongest two-sided laboratory bounds on $c_{TT}$, and the
radiation spectrum is sensitive to rotation noninvariance coefficients as well.
Synchrotron motion and radiation in the presence of $c$ was discussed
in~\cite{ref-altschul5}. The basic picture is fairly simple, and the instantaneous
radiation rate depends on the difference between an orbiting charge's speed and the
speed of light in the direction the charge is moving.

With $k_{F}=0$,
the electromagnetic sector behaves according to ordinary special relativity; the
electromagnetic field responds to charged particles normally, and the radiation
emitted is determined in the usual way by the velocity profiles of the moving
charges. The total radiated synchrotron power is
$P=\frac{e^{2}a^{2}}{6\pi m^{2}}\gamma^{4}$,
where $a$ is the magnitude of the magnetic acceleration. [If photon-sector
Lorentz violation were included, the electromagnetic sector would behave (in the
narrow pencil of angles into which the synchrotron radiation is beamed) like standard
special relativity, except with a modified Lorentz factor
\begin{equation}
\tilde{\gamma}=\frac{1}{\sqrt{1+2\delta_{\gamma}(\hat{v})-v^{2}}}.
\end{equation}
The power radiated would then be
$P=\frac{e^{2}a^{2}}{6\pi m^{2}}\tilde{\gamma}^{4}$.]
For ul\-tra\-rel\-a\-tiv\-ist\-ic par\-ti\-cles, $\gamma\approx[2(1-v)]^{-1/2}$
increases very rapidly as a function of $v$, since
$\frac{d\gamma}{dv}=v\gamma^{3}\approx\gamma^{3}$. The modified expression for
$\vec{v}(\vec{p}\,)$ changes the radiated power $P(\vec{p}\,)$ to
\begin{equation}
\label{eq-P}
P(\vec{p}\,)=P_{0}(\vec{p}\,)\left\{1+4\gamma^{2}\left[\delta(\hat{p})-
\delta_{\gamma}(\hat{p})\right]\right\},
\end{equation}
where $P_{0}(\vec{p}\,)$ is the radiation rate for a particle of momentum
$\vec{p}$ in the
absence of any Lorentz violation. The fractional change in the radiated power
increases with the square of the energy.

The beam energy at LEP was determined very precisely, because making high-precision
measurements of the W and Z boson masses was one of the accelerator's most important
functions.
The beam energy $E_{b}$ was calculated using several different methods, which could
be used to double-check one-another.
The primary method was based on measuring the magnetic field profile along the
beam's trajectory (which was also carefully measured). The field measurements
used nuclear magnetic resonance (NMR) probes.
The validity of the NMR measurements should not be affected by Lorentz violation;
Lorentz violation strong enough to affect the NMR is already ruled out by
atomic clock experiments~\cite{ref-berglund,ref-kost6,ref-bear,ref-wolf}.
The field profile and the trajectory through the bending magnets were known to high
precision, and together these determined $E_{b}$.
 
A complementary measurement used the  synchrotron tune, $Q_{s}$---the ratio of the
synchrotron oscillation frequency to the
orbital frequency~\cite{ref-assmann}. Synchrotron oscillations occur because of
nonuniformity in beam particles' energies. Particles with less than the
nominal beam energy curve more sharply and thus travel around smaller orbits. This
means they take less time to travel between the accelerating radio frequency (RF)
cavities. The less energetic particles therefore arrive at the cavities earlier in
the RF cycle and receive larger-than-expected energy boosts, moving their energy back
toward the nominal beam energy. An opposite effect occurs for
particle with greater than the nominal energy. These effects cause
oscillations in the energy of the beam, and a fit to their frequency provides an
independent way to determine the central energy $E_{b}$.
The best measurement of the synchrotron tune at LEP came from runs at $E_{b}=91$ GeV,
near the Z pole. The Lorentz factor at this energy is
$\gamma>1.7\times 10^{5}$.
Fitting the synchrotron oscillations at this value of the beam energy, the difference
between the values of the energy inferred from
the tune $Q_{s}$ and from the magnetic field
profile was only 3 MeV; this discrepancy was small compared with the $1\sigma$
uncertainty of 21 MeV that was associated with the fitting procedure and with the
imprecision in the actual measurement of $Q_{s}$.

The formula for $Q_{s}$ depends on $E_{b}$ and the energy loss per orbit $U_{0}$, in
the combination $(g^{2}e^{2}V_{RF}^{2}-U_{0}^{2})/E_{b}^{2}$.
$V_{RF}$ is the amplitude of the RF voltage during the beam's passage through the
accelerating cavities; $\psi_{s}$ is the phase of the accelerating voltage, and under
the steady state conditions at which the collider operated,
$U_{0}=geV_{RF}\sin\psi_{s}$. Both $V_{RF}$ and $\psi_{s}$ are known to high
precision. The factor $g$ is a correction associated with possible phasing errors and
misalignments of the RF cavities; $g$ must be determined as part of the fit for
$Q_{s}$, although its value is necessarily close to 1. The fitting required
makes $g$ a significant source of uncertainty. However, the fitting can
be done on data taken at well below the usual experimental energies, since the magnet
alignment and RF cavity timing do not depend on the operating energy. At low
energies,
$g$ can be determined more precisely than would be possible at higher energies; and
the fitted value of the correction factor can then be carried over to the actual 91
GeV running energy.
The possible presence of Lorentz violation changes the formula for $Q_{s}$ only
through a rescaling of $U_{0}$; the Lorentz force law is not modified, and the
energy imparted by the RF cavities is unchanged.

In the fit used to determine $Q_{s}$, it was assumed that $U_{0}$ had the usual
synchrotron form plus small corrections. The corrections are necessary to account for
other forms of energy loss, primarily through parasitic impedance interactions
between the beam and the apparatus or effects associated with finite beam size.
However, these corrections could be either modeled from first principles or measured
directly. Since the NMR measurements provided an independent measure of the
beam energy $E_{b}$, it is possible to reinterpret the fit for $Q_{s}$ as a
measurement not of $E_{b}$ but rather of $U_{0}$. Because $E_{b}$ and $U_{0}$ enter
the formula for $Q_{s}$ primarily in the combination $(U_{0}/E_{b})^{2}\cos^{2}\psi_{s}$, if the fit (assuming a known form for $U_{0}$)
could determine $E_{b}$ with a fractional uncertainty $\eta$, the same fit would
(if $E_{b}$ were instead taken to be known) determine $U_{0}$ with the same
fractional precision. Since $U_{0}$ is determined by the synchrotron radiation power
$P$, this immediately places a bound on the SME contribution to $P$ as given in
(\ref{eq-P}):
\begin{equation}
\label{eq-deltaave}
|\langle\delta\rangle|<C\equiv\eta\frac{m^{2}}{4E_{b}^{2}},
\end{equation}
where $\langle\delta\rangle$ denotes the
average value of $\delta(\hat{p})$ over one orbit around
the accelerator.

In $\langle\delta\rangle$, the parity-odd $c_{0j}$ coefficients average to
zero. The effects of these coefficients on the speed $v$ change sign when the
momentum is reversed; they cannot contribute to the average radiation emitted over
a full revolution, because orbiting particles experience equal and opposite
velocities at antipodal points along their orbits. The effects of the $d$
coefficients are also averaged out. The $d$ coefficients would affect the speed and
hence the synchrotron losses if the beams were consistently longitudinally polarized.
However, such polarization cannot be maintained over long periods; helicities evolve
in the magnetic field, because of particles' anomalous magnetic moments. The beams at
LEP were maintained through most of their orbits in transverse polarization states;
if desired, they could be temporarily rotated into longitudinal polarization states
immediately before an interaction point. The average value of $\delta(\hat{p})$
is thus $-c_{00}-\frac{1}{2}c_{j}c_{k}\left[(\hat{e}_{1})_{j}(\hat{e}_{1})_{k}+
(\hat{e}_{2})_{j}(\hat{e}_{2})_{k}\right]$,
where $\hat{e}_{1}$ and $\hat{e}_{2}$ are perpendicular unit vectors in the plane
of the orbit; a particle essentially performs half its motion in the
$\pm\hat{e}_{1}$-direction and half in the $\pm\hat{e}_{2}$-direction. Using the
tracelessness of $c^{\nu\mu}$, the average may be more concisely written as
\begin{equation}
\label{eq-deltan}
\langle\delta\rangle=-\frac{3}{2}c_{00}+\frac{1}{2}c_{jk}\hat{n}_{j}\hat{n}_{k},
\end{equation}
where $\hat{n}$ is the unit vector normal to the circular path.

The fractional uncertainty assigned to the value of $E_{b}$ that was derived from
fitting the synchrotron tune was $\eta=2.4\times 10^{-4}$. This could
then be interpreted as the fractional error in $U_{0}$ and hence $P$.
A somewhat more conservative estimate of the possible error is
$\eta <6\times 10^{-4}$; this is a $2\sigma$ error, and it also accounts for other
sources of error, such as in the NMR measurement of $E_{b}$ and the discrepancy
between the NMR and $Q_{s}$ measurements of the energy. Using this value for $\eta$,
we have $C=5\times10^{-15}$.

This provides a bound on a combination of boost and rotation invariance violation
coefficients. The relevant $\hat{n}$ appearing in (\ref{eq-deltan}) is the normal
vector to the orbit at the time
when the synchrotron tune data used in the energy calibration was collected. If the
tune data was collected over an extended period (covering many values of $\hat{n}$),
and $Q_{s}$ was consistent across
all these measurements, the bound (\ref{eq-deltaave})
would hold for all the directions $\hat{n}$ that the accelerator passed through---all
$\hat{n}$ with $\hat{n}\cdot\hat{Z}=\cos\chi_{C}$. However, this conclusion
does not appear to be supported; it is only the average value of the synchrotron tune
that was used for the energy fitting, and we cannot rule out that there were small
sidereal variations in $Q_{s}$. We could use the sidereally averaged value
$\langle c_{jk}\hat{n}_{j}\hat{n}_{k}\rangle=\frac{1}{2}\sin^{2}\chi_{C}(c_{XX}+
c_{YY})+\cos^{2}\chi_{C}\,c_{ZZ}$, but there are other, stronger laboratory bounds
on the $c_{JK}$ coefficients, which will described in section~\ref{sec-comb}; and
the existence of these $c_{JK}$ bounds makes (\ref{eq-deltaave}) important primarily
as a bound on $c_{TT}$.

\section{Best Combined Bounds on $c^{\nu\mu}$}
\label{sec-comb}

We shall finally turn our attention to how the high-energy bounds (and other,
low-energy bounds to be described shortly) can be merged together to produce the
best laboratory constraints on the $c^{\nu\mu}$ coefficients. We shall pay
particular attention to those coefficients, $c_{TT}$ and $c_{(TJ)}$ associated with
the breaking of boost invariance. (The $d^{\nu\mu}$ coefficients, as previously
noted, are already extremely well constrained by nonrelativistic spin measurements.)

The synchrotron bounds take the form
$\left|\frac{3}{2}c_{TT}-\frac{1}{2}c_{JK}\hat{n}_{J}\hat{n}_{K}\right|<C$.
Regardless of what direction $\hat{n}$ is used, this implies that
$|c_{TT}|<\frac{3}{4}(C+\lambda)$, where $\lambda$ is the largest singular value of
the traceless $3\times3$ matrix
\begin{equation}
c_{JK}-\frac{1}{3}c_{TT}\delta_{JK}=\frac{1}{2}\left[
\begin{array}{ccc}
c_{-}+\frac{1}{3}c_{Q} & c_{(XY)} & c_{(XZ)} \\
c_{(XY)} & -c_{-}+\frac{1}{3}c_{Q} & c_{(YZ)} \\
c_{(XZ)} &  c_{(YZ)} & -\frac{2}{3}c_{Q}
\end{array}
\right].
\end{equation}
The quantities $c_{-}=c_{XX}-c_{YY}$, $c_{Q}=c_{XX}+c_{YY}-2c_{ZZ}$, and $c_{(JK)}$
are those that can be bounded in stationary tests of isotropy. The singular value
must satisfy
\begin{equation}
\label{eq-lambda}
\lambda\leq\frac{1}{\sqrt{2}}\left[|c_{(XY)}|^{2}+|c_{(XZ)}|^{2}+|c_{(YZ)}|^{2}+
|c_{-}|^{2}+\frac{1}{3}|c_{Q}|^{2}\right]^{1/2}.
\end{equation}
The existence of multiple complementary bounds on the types of Lorentz
violation~\cite{ref-herrmann2,ref-herrmann3,ref-eisele}
relevant to optical cavity experiments should imply bounds on the $c$ coefficients
appearing in (\ref{eq-lambda}), each at the $10^{-17}$ level.
The key point in the derivation of such electron bounds from Michelson-Morley-type
measurements
%of the isotropy of photon propagation in cavities
is the use of
optical cavities containing different materials.
The dimensions of a block of material determine its resonant frequencies when it
is used as an optical cavity. However, these dimensions are determined by the
material's electron structure and are affected by electron Lorentz violation.
This makes the Michelson-Morley experiments sensitive to
both photon and electron sector SME coefficients. Cavity materials with different
electronic properties are sensitive to different mixtures of electron and photon
parameters, so with multiple experiments, it is possible to disentangle the bounds
from the two sectors~\cite{ref-muller2}. 

An up-to-date analysis of these bounds has not been performed. Older
resonator data provided bounds on all the $c$ coefficients appearing in
(\ref{eq-lambda}) at the $10^{-16}$ level, except for $c_{Q}$, which was bounded at
only the $10^{-14}$ level~\cite{ref-muller3}. However, we shall assume that an
improved analysis based on~\cite{ref-herrmann2,ref-herrmann3,ref-eisele}
will indeed produce bounds on the five anisotropic $c_{JK}$ coefficients (and hence
$\lambda$) at approximately
the $10^{-17}$ level. This means that these coefficients can effectively
be neglected in the analysis of the accelerator results, for purposes of determining
the strongest laboratory bounds that are available on $c_{TT}$; $\lambda$ is very
small compared to $C$. This implies that
\begin{equation}
\label{eq-sync-cTTbound}
|c_{TT}|<4\times10^{-15};
\end{equation}
this is
the same bound given in~\cite{ref-altschul20}, but expressed in different notation,
with the photon sector Lorentz violation set to zero.

The synchrotron bound on $c_{TT}$, the resonant cavity bounds on the anisotropic
$c_{JK}$, the torsion pendulum bounds on the $d^{\nu\mu}$, and the vacuum Cerenkov
bounds (\ref{eq-vcbound}) can be further combined to establish
new, accelerator-based constraints on the $c_{(TJ)}$ parameters. Because of the
smallness of the $|c_{JK}|\lesssim\lambda$ and $|c_{TT}|\lesssim C$, in conjunction
with the hierarchy $\lambda\ll C\ll A$, the $\delta(\hat{v})$ appearing on the
left-hand side of (\ref{eq-vcbound}) will be dominated by $\pm v_{J}c_{(TJ)}$ if the
Lorentz violation is at the greatest allowed level.
The corresponding bounds are (neglecting $d$ entirely)
$|v_{J}c_{(TJ)}|<A+C+2\lambda$, or
\begin{eqnarray}
\label{eq-cTJbound}
|c_{(TX)}|,\,|c_{(TY)}| & < & 1.2\times10^{-11} \\
\label{eq-cTZbound}
|c_{(TZ)}| & < & 1.8\times10^{-11}.
\end{eqnarray}
These are an order of magnitude better than the best atomic clock bounds on these
kinds of parameters~\cite{ref-altschul21}.

However, since the reliability of a bound based on the absence of
$\gamma\rightarrow e^{+}+e^{-}$ does not depend on how the $\gamma$-ray involved
originated, we can improve these bounds using astrophysical observations.
In~\cite{ref-altschul14}, we discussed the bounds that could be placed on
$c$ using astrophysical photon survival data and compiled a list of sources that
produce $\gamma$-rays with $E_{\gamma}/m_{e}$ in the $10^{7}$--$10^{8}$ range.
To each such source, there corresponds a constraint of the form
(\ref{eq-photonbound}), parameterized by the direction $\hat{v}$ from the source to
the Earth.
In terms of the right ascension $\alpha$ and declination $\delta$, the components
of this $\hat{v}$ are $\hat{v}_{X}=-\cos\delta\cos\alpha$,
$\hat{v}_{Y}=-\cos\delta\sin\alpha$, and $\hat{v}_{Z}=-\sin\delta$. Using only the
photon survival data, it was not possible to place two-sided constraints.
However, two-sided bounds on the $c_{(TJ)}$ are again possible if the photon
survival data is combined with the same laboratory bounds on $c_{TT}$ and
$c_{JK}$ that were discussed in the preceding paragraph.

\begin{table}
\begin{center}
\begin{tabular}{|l|c|c|c|c|}
\hline
Emission source & $\hat{v}_{X}$ & $\hat{v}_{Y}$ & $\hat{v}_{Z}$ &
$E_{\gamma}/m_{e}$ \\
\hline
Cas A & $-0.51$ & 0.08 & $-0.86$ &
$8\times 10^{6}$\cite{ref-acciari1} \\
Crab nebula & $-0.10$ & $-0.92$ & $-0.37$ 
& $1.6\times 10^{8}$\cite{ref-tanimori} \\
G 12.82-0.02 & $-0.06$ & 0.95 & 0.29 &
$5\times 10^{7}$\cite{ref-aharonian6} \\
G 18.0-0.7 & $-0.11$ & 0.97 & 0.24 &
$7\times 10^{7}$\cite{ref-aharonian8,ref-aharonian11} \\
G 106.3+2.7 & $-0.45$ & 0.19 & $-0.87$ &
$2\times 10^{7}$\cite{ref-acciari2} \\
G 347.3-0.5 & 0.16 & 0.75 & 0.64 &
$10^{8}$\cite{ref-aharonian15} \\
J1427-608 & 0.39 & 0.29 & 0.87 &
$6\times10^{7}$\cite{ref-aharonian13} \\
J1626-490 & 0.26 & 0.60 & 0.75 &
$6\times10^{7}$\cite{ref-aharonian13} \\
J1731-437 & 0.09 & 0.72 & 0.69 &
$6\times10^{7}$\cite{ref-aharonian13} \\
J1745-303 & 0.06 & 0.86 &  0.50 &
$4\times10^{7}$\cite{ref-aharonian14} \\
J1841-055 & $-0.18$ & 0.98 & 0.10 &
$4\times10^{7}$\cite{ref-aharonian13} \\
J1857+026 & $-0.25$ & 0.97 & $-0.05$ &
$8\times10^{7}$\cite{ref-aharonian13} \\
M87 & 0.97 & 0.13 & $-0.21$ &
$2\times10^{7}$\cite{ref-aharonian12} \\
Mkn 421 & 0.76  & $-0.19$  & $-0.62$ &
$3\times 10^{7}$\cite{ref-albert,ref-aharonian7} \\
Mkn 501 & 0.22 &0.74 & $-0.64$ &
$4\times 10^{7}$\cite{ref-aharonian10} \\
MSH 15-52 & 0.34 & 0.38 & 0.86 &
$8\times 10^{7}$\cite{ref-aharonian4} \\
Vela SNR & 0.44 & $-0.55$ & 0.71 &
$1.3\times 10^{8}$\cite{ref-aharonian2} \\
\hline
\end{tabular}
\caption{
\label{table-Eobs}
Energies of observed $\gamma$-rays from various astrophysical sources. References
are given for each value of the energy.}
\end{center}
\end{table}

Table~\ref{table-Eobs} gives an updated list of sources for which TeV photon
observations have been made. (This list is not exhaustive. There are other sources
from which comparably energetic photons have been observed, but they are not included
if the do not affect the output of the linear program discussed below. A source may
be superfluous in this way if, for example, it is located quite close to
another source, so that the two sources produce bounds
on very similar linear combinations of coefficients.)
Most of the best measurements of ultra-high-energy photon
spectra have been made by the H.E.S.S. atmospheric Cerenkov telescope in Namibia
and the newer VERITAS telescope in Arizona.
These devices are extremely sensitive, but they have limited sky coverage.
%which forces us to rely on accelerator data to a greater extent that would be ideal.

To separate the $c_{(TJ)}$ bounds from those on other coefficients
requires linear programming, but the results are
straightforward. Using the the data from table~\ref{table-Eobs} in the
inequality~(\ref{eq-photonbound}), as well as using
the inequality (\ref{eq-sync-cTTbound}) [inclusion of the
LEP photon survival results, in the form of inequalities (\ref{eq-cTJbound}) and
(\ref{eq-cTZbound}), turns out to be unnecessary], we have
extracted bounds on all the boost-invariance-violating $c$ coefficients. Because the
resulting bounds are significantly less stringent than the expected
Michelson-Morley bounds on the $c_{JK}$, we can neglect the $c_{JK}$
entirely in our analysis.

\begin{table}
\begin{center}
\begin{tabular}{|c|c|c|}
\hline
$c^{\mu\nu}$ & Maximum & Minimum \\
\hline
$c_{TT}$ & $2\times 10^{-15}$ & $-4\times 10^{-15}$ \\
$c_{(TX)}$ & $10^{-14}$ & $-3\times 10^{-13}$ \\
$c_{(TY)}$ & $6\times 10^{-15}$ & $-8\times 10^{-14}$ \\
$c_{(TZ)}$ & $1.3\times 10^{-13}$ & $-1.1\times 10^{-14}$ \\
\hline
\end{tabular}
\caption{
\label{table-bounds}
Disentangled bounds on the boost invariance violation components of $c$.}
\end{center}
\end{table}

The results from the linear program are given in table~\ref{table-bounds}. The values
listed in the table are the absolute maximum and minimum values of the $c_{TT}$ and
$c_{(TJ)}$ coefficients that are consistent with the data. No restrictions on any
other coefficients are assumed, except with regard to a convention for choosing
coordinates.
Our convention has been to set $\tilde{k}^{\mu\nu}=0$. If the coordinates are
not chosen so as to make the photon sector conventional, all these bounds are really
on linear combinations
$c^{\nu\mu}-\frac{1}{2}\tilde{k}^{\mu\nu}$. In terms of the $\tilde{\kappa}$
three-tensors into which $k_{F}$ is usually decomposed, the bounds are on the
combinations $c_{TT}-\frac{3}{4}\tilde{\kappa}_{{\rm tr}}$ and
$c_{(TJ)}-\frac{1}{2}\epsilon_{JKL}(\tilde{\kappa}_{o+})_{KL}$.
None of the bounds in table~\ref{table-bounds} are as strong as the bounds on the
same quantities from~\cite{ref-altschul7}, which were derived by assuming we possess
an accurate
understanding of how the sources involved are producing their spectra. However, for
the upper bound on $c_{(TY)}$, the difference in precision is less than an order of
magnitude.

Which
physical phenomena are responsible for which of the bounds deserves some explanation.
The lower bound on $c_{TT}$ is simply that of (\ref{eq-sync-cTTbound}), but the
upper bound also depends on both H.E.S.S. and VERITAS astrophysical photon survival
data. The upper bounds on $c_{(TX)}$ and $c_{(TY)}$ and the lower bound on $c_{(TZ)}$
come from combining the H.E.S.S. data with the synchrotron bounds on $c_{TT}$.
The remaining three bounds---the lower bounds on $c_{(TX)}$ and $c_{(TY)}$ and the
upper bound on $c_{(TZ)}$---are roughly an order of magnitude weaker. These bounds
depend on the inclusion of VERITAS data, which are not yet as comprehensive as
the H.E.S.S. data at the very highest energies. Yet without the the VERITAS
observations, the bounds in question would be at the $10^{-11}$ level corresponding
to (\ref{eq-cTJbound}) and (\ref{eq-cTZbound}).

It is also important to understand the advantages and limitations of each of
the methods of bounding $c_{TT}$ and $c_{(TJ)}$.
The strength of the $c_{TT}$ bounds is a consequence of the quality of the LEP
energy calibration data. The analysis of the of the synchrotron spectrum
gives bounds that are proportional to $\frac{m^{2}}{E^{2}}$; the same factor
determines the strength of the threshold bounds. However, the bounds involving
thresholds for new processes cannot be improved by more precise measurements;
if a decay process like vacuum Cerenkov radiation is allowed at a certain
energy, it will generally occur quite quickly and be easily detectable. If no
such decays are seen to occur, a bound on the SME coefficients immediately
follows. The difference with synchrotron radiation is that it occurs whether
the theory is conventional or Lorentz violating. This means the spectrum can be
measured precisely, and the precision factor $\eta$ improves the bounds on the
$c$ coefficients involved. In principle, the
instantaneous synchrotron spectrum is sensitive to
all the electron $c$ coefficients. However, the total emission over one orbit is
the only quantity that has been accurately determined, and this is sensitive
only to a particular combination of coefficients---$\langle\delta\rangle$ from
(\ref{eq-deltaave})---which does not involve the $c_{(TJ)}$.

The observed absences of
$e^{\pm}\rightarrow e^{\pm}+\gamma$ and $\gamma\rightarrow e^{+}+e^{-}$ up to
an energy $E$ also constrain $c_{TT}$ and $c_{(TJ)}$ at the
${\cal O}\left(\frac{m^{2}}{E^{2}}\right)$
level. The LEP data show the absence of vacuum Cerenkov radiation up to beam
energies of 104.5 GeV. This is a much lower energy than can be seen in cosmic
ray photons; the raw bounds (\ref{eq-photonbound}) derived from the H.E.S.S.
observations are 4--5 orders of magnitude stronger than the LEP bounds
(\ref{eq-vcbound}). However, the advantage of the accelerator data is that it
is available and consistent over quite a broad range of directions $\hat{v}$.
Until the publication of the VERITAS data, the H.E.S.S. data alone were insufficient
to constrain all the $c_{(TJ)}$ coefficients from above and below; the LEP results
would have been needed to fill in some gaps.

Now, with VERITAS accumulating data and covering a quite different area of the sky
from H.E.S.S., those LEP results are no longer needed in order to set reliable
laboratory bounds on the $c_{(TJ)}$. Naturally, as both $\gamma$-ray telescopes
explore more sources for longer times, the quality of the bounds will improve
somewhat. It should be possible to improve the laboratory
bounds on all the $c_{(TJ)}$ to the $10^{-14}$ level or better.

Of course, in deriving the bounds on $c_{TT}$ and $c_{(TJ)}$, we have assumed that
the Mi\-chel\-son-Morley experiments do indeed place strong enough bounds on all the
stationary anisotropy coefficients so that $\lambda$ may effectively be neglected.
Existing analyses have not quite achieved this, although the raw Michelson-Morley
data certainly appear to be sufficiently precise. Moreover,
of the five coefficients that can, according to (\ref{eq-lambda}), contribute to
$\lambda$, only $c_{Q}$ is potentially large enough to interfere with the
results in this paper. It would be quite desirable to  have updated bounds on the
$c_{JK}$ coefficients, based on the latest generation of resonant cavity
experiments. Such bounds would be interesting, both for their influence on the
effects discussed here and in their own right. However, if improved bounds on
$c_{Q}$ do not actually materialize, the effect on the results of this paper
would be modest. None of the bounds in table~\ref{table-bounds} would be worsened
by more than $4\times10^{-14}$.

We have discussed several different phenomena that can be observed in the laboratory
and used to constrain electron Lorentz violation.
However, not all of these methods produce
competitive bounds. The best present laboratory
techniques for bounding the various SME coefficients we have
discussed are:  torsion pendulum experiments with spin-polarized samples for
$d^{\nu\mu}$, Michelson-Morley experiments with resonators of different compositions
for $c_{JK}$, synchrotron radiation at LEP for $c_{TT}$, and observations of photon
survival for $c_{(TJ)}$. The last of these may involve $\gamma$-rays of astrophysical
origin, but which are measured in a laboratory on Earth. Using all the observations of
such $\gamma$-rays, we have placed new laboratory bounds on all the $c_{(TJ)}$
coefficients, at the $10^{-13}$--$10^{-15}$ levels.
[Using more restricted collections of data sets would lead to results such as
(\ref{eq-cTTbound}--\ref{eq-dTJbound}) or (\ref{eq-cTJbound}--\ref{eq-cTZbound}).]
Some of these bounds are
naturally expected to improve as better observational data become available.

%\section*{Acknowledgments}
%The author is grateful to XXX for helpful discussion.


\begin{thebibliography}{99}


\bibitem{ref-kost1}D. Colladay, V. A. Kosteleck\'{y}, Phys. Rev. D {\bf 55},
6760 (1997).
\bibitem{ref-kost2}D. Colladay, V. A. Kosteleck\'{y}, Phys. Rev. D {\bf 58},
116002 (1998).
\bibitem{ref-kost12}V. A. Kosteleck\'{y}, Phys. Rev. D, {\bf 69} 105009 (2004).
\bibitem{ref-kost4}V. A. Kosteleck\'{y}, C. D. Lane, A. G. M. Pickering,
Phys. Rev. D {\bf 65}, 056006 (2002).
\bibitem{ref-colladay2}D. Colladay, P. McDonald, Phys. Rev. D {\bf 75}, 105002
(2007).
\bibitem{ref-kost3}V. A. Kosteleck\'{y}, R. Lehnert, Phys. Rev. D {\bf 63},
065008 (2001).
\bibitem{ref-bluhm1}R. Bluhm, V. A. Kosteleck\'{y}, N. Russell, Phys. Rev.
Lett. {\bf 79}, 1432 (1997).
\bibitem{ref-gabirelse}G. Gabrielse, A. Khabbaz, D. S. Hall, C. Heimann, H.
Kalinowsky, W. Jhe, Phys. Rev. Lett. {\bf 82}, 3198 (1999).
\bibitem{ref-dehmelt1}H. Dehmelt, R. Mittleman, R. S. Van Dyck, Jr., P.
Schwinberg, Phys. Rev. Lett. {\bf 83}, 4694 (1999).
\bibitem{ref-bluhm3}R. Bluhm, V. A. Kosteleck\'{y}, N. Russell , Phys. Rev.
Lett. {\bf 82}, 2254 (1999).
\bibitem{ref-phillips}D. F. Phillips, M. A. Humphrey, E. M. Mattison, R. E.
Stoner, R. F. C. Vessot, R. L. Walsworth , Phys. Rev. D {\bf 63}, 111101 (R)
(2001).
\bibitem{ref-kost8}R. Bluhm, V. A. Kosteleck\'{y}, C. D. Lane, Phys. Rev. Lett.
{\bf 84}, 1098 (2000).
\bibitem{ref-hughes}V. W. Hughes, {\em et al.}, Phys. Rev. Lett. {\bf 87},
111804 (2001).
\bibitem{ref-heckel3}B. R. Heckel, E. G. Adelberger, C. E. Cramer, T. S. Cook, S.
Schlamminger, U. Schmidt, Phys. Rev. D {\bf 78}, 092006 (2008).
\bibitem{ref-berglund}C. J. Berglund, L. R. Hunter, D. Krause, Jr., E. O.
Prigge, M. S. Ronfeldt, S. K. Lamoreaux, Phys. Rev. Lett. {\bf 75}, 1879 (1995).
\bibitem{ref-kost6}V. A. Kosteleck\'{y}, C. D. Lane, Phys. Rev. D {\bf 60},
116010 (1999).
\bibitem{ref-bear}D. Bear, R. E. Stoner, R. L. Walsworth, V. A. Kosteleck\'{y},
C. D. Lane, Phys. Rev. Lett. {\bf 85}, 5038 (2000).
\bibitem{ref-wolf}P. Wolf, F. Chapelet, S. Bize, A. Clairon,  Phys. Rev. Lett.
{\bf 96}, 060801 (2006).
\bibitem{ref-muller3}H. M\"{u}ller, {\em et al.}, Phys. Rev. Lett. {\bf 99}, 050401
(2007).
\bibitem{ref-herrmann2}S. Herrmann, A. Senger, K. M\"{o}hle, E. V. Kovalchuk, A.
Peters, in {\em CPT and Lorentz Symmetry IV}, edited by V. A. Kosteleck\'{y} (World
Scientific, Singapore, 2008), p. 9.
\bibitem{ref-herrmann3}S. Herrmann, {\em et al.}, Phys. Rev. D 80, 105011 (2009).
\bibitem{ref-eisele}Ch. Eisele, A. Yu. Nevsky, S. Schiller, Phys. Rev.
Lett. 103, 090401 (2009).
\bibitem{ref-muller2}H. M\"{u}ller, Phys. Rev. D {\bf 71}, 045004 (2005).
\bibitem{ref-saathoff}G. Saathoff, S. Karpuk, U. Eisenbarth, G. Huber, S. Krohn, R.
Mu\~{n}oz Horta, S. Reinhardt, D. Schwalm, A. Wolf, G. Gwinner, Phys. Rev. Lett.
{\bf 91}, 190403 (2003).
\bibitem{ref-lane1}C. D. Lane, Phys. Rev. D {\bf 72}, 016005 (2005).
\bibitem{ref-kost10}V. A. Kosteleck\'{y}, Phys. Rev. Lett. {\bf 80}, 1818
(1998).
\bibitem{ref-kost7}V. A. Kosteleck\'{y}, Phys. Rev. D {\bf 61}, 016002 (1999).
\bibitem{ref-hsiung}Y. B. Hsiung, Nucl. Phys. Proc. Suppl. {\bf 86}, 312
(2000).
\bibitem{ref-abe}K. Abe {\em et al.}, Phys. Rev. Lett. {\bf 86}, 3228 (2001).
\bibitem{ref-link}J. M. Link {\em et al.}, Phys. Lett. B {\bf 556}, 7 (2003). 
\bibitem{ref-aubert}B. Aubert {\em et al.}, Phys. Rev. Lett. {\bf 96}, 251802 (2006).
\bibitem{ref-carroll2}S. M. Carroll, G. B. Field, Phys. Rev. Lett. {\bf 79},
2394 (1997).
\bibitem{ref-kost11}V. A. Kosteleck\'{y}, M. Mewes, Phys. Rev. Lett. {\bf 87},
251304 (2001).
\bibitem{ref-kost21}V. A. Kosteleck\'{y}, M. Mewes, Phys. Rev. Lett. {\bf 97},
140401 (2006).
\bibitem{ref-kost22}V. A. Kosteleck\'{y}, M. Mewes, Phys. Rev. Lett. {\bf 99}, 011601
(2007).
\bibitem{ref-stecker}F. W.  Stecker, S. L.  Glashow, Astropart. Phys. {\bf 16},  97
(2001).
\bibitem{ref-jacobson1}T. Jacobson, S. Liberati, D. Mattingly, Nature {\bf 424},
1019 (2003).
\bibitem{ref-altschul6}B. Altschul, Phys. Rev. Lett. {\bf 96}, 201101 (2006).
\bibitem{ref-altschul7}B. Altschul, Phys. Rev. D {\bf 74}, 083003 (2006).
\bibitem{ref-klinkhamer2}F. R. Klinkhamer, M. Risse, Phys. Rev. D {\bf 77}, 016002
(2008); addendum Phys. Rev. D {\bf 77}, 117901 (2008).
\bibitem{ref-battat}J. B. R. Battat, J. F. Chandler, C. W. Stubbs, Phys. Rev. Lett.
{\bf 99}, 241103 (2007).
\bibitem{ref-muller4}H. M\"{u}ller, S. Chiow, S. Herrmann, S. Chu, Phys. Rev. Lett.
{\bf 100}, 031101 (2008).
\bibitem{ref-tables}V. A. Kosteleck\'{y}, N. Russell, arXiv:0801.0287.
\bibitem{ref-altschul4}B. Altschul, D. Colladay, Phys. Rev. D {\bf 71}, 125015
(2005).
\bibitem{ref-bluhm4}R. Bluhm, V. A. Kosteleck\'{y}, C. D. Lane, N. Russell, Phys.
Rev. D {\bf 68}, 125008 (2003).
\bibitem{ref-altschul9}B. Altschul, Phys. Rev. Lett. {\bf 98}, 041603 (2007).
\bibitem{ref-hohensee1}M. A. Hohensee, R. Lehnert, D. F. Phillips, R. L. Walsworth,
Phys. Rev. Lett. {\bf 102}, 170402 (2009).
\bibitem{ref-coleman2}S. Coleman, S. L. Glashow, Phys. Rev. D {\bf 59}, 116008
(1999).
\bibitem{ref-hohensee2}M. A. Hohensee, R. Lehnert, D. F. Phillips, R. L. Walsworth,
Phys. Rev. D {\bf 80}, 036010 (2009). 
\bibitem{ref-abazov}V. M. Abazov, {\em et al.}, Phys. Lett. B {\bf 639}, 151 (2006).
\bibitem{ref-altschul5}B. Altschul, Phys. Rev. D {\bf 72}, 085003 (2005).
\bibitem{ref-assmann}R. Assmann, {\em et al.}, Eur. Phys. J. C 39, {\bf 253} (2005).
\bibitem{ref-altschul20}B. Altschul, Phys. Rev. D {\bf 80}, 091901(R) (2009).
\bibitem{ref-altschul21}B. Altschul, Phys. Rev. D {\bf 81}, 041701(R) (2010).
\bibitem{ref-altschul14}B. Altschul, Astropart. Phys. {\bf 28}, 380 (2007).
\bibitem{ref-acciari1}V. A. Acciari, {\em et al.}, Astrophys. J. {\bf 714}, 163
(2010).
\bibitem{ref-tanimori}T. Tanimori, {\em et al.}, Astrophys. J. Lett. {\bf 492}, L33
(1998).
\bibitem{ref-aharonian6}F. A. Aharonian, {\em et al.}, Astrophys. J. {\bf 636}, 777
(2006).
\bibitem{ref-aharonian8}F. A. Aharonian, {\em et al.}, Astron. Astrophys. {\bf 442},
L25 (2005).
\bibitem{ref-aharonian11}F. A. Aharonian, {\em et al.}, astro-ph/0607548.
\bibitem{ref-acciari2}V. A. Acciari, {\em et al.}, Astrophys. J. Lett. {\bf 703}, L6
(2009).
\bibitem{ref-aharonian15}F. A. Aharonian, {\em et al.}, Astron. Astrophys. {\bf 464},
235 (2007).
\bibitem{ref-aharonian13}F. A. Aharonian, {\em et al.}, Astron. Astrophys. {\bf 477},
353 (2008).
\bibitem{ref-aharonian14}F. A. Aharonian, {\em et al.}, Astron. Astrophys. {\bf 483},
509 (2008).
\bibitem{ref-aharonian12}F. A. Aharonian, {\em et al.}, Science {\bf 314}, 1424
(2006).
\bibitem{ref-albert}J. Albert, {\em et al.}, Astrophys. J. {\bf 663}, 125 (2007).
\bibitem{ref-aharonian7}F. A. Aharonian, {\em et al.}, Astron. Astrophys. {\bf 437},
95 (2005).
\bibitem{ref-aharonian10}F. A. Aharonian, {\em et al.}, Astron. Astrophys. {\bf 349},
11 (1999).
\bibitem{ref-aharonian4}F. A. Aharonian, {\em et al.}, Astron. Astrophys. {\bf 435},
L17 (2005).
\bibitem{ref-aharonian2}F. A. Aharonian, {\em et al.}, Astron. Astrophys. {\bf 448},
L43 (2006).

\end{thebibliography}
\end{document}